\documentclass[fleqn,usenatbib]{mnras}

\usepackage{newtxtext,newtxmath}

\usepackage[T1]{fontenc}

\DeclareRobustCommand{\VAN}[3]{#2}
\let\VANthebibliography\thebibliography
\def\thebibliography{\DeclareRobustCommand{\VAN}[3]{##3}\VANthebibliography}

\usepackage[normalem]{ulem}
\usepackage{tabularx}
\useunder{\uline}{\ul}{}
\usepackage{xcolor}
\usepackage{graphicx}	
\usepackage{amsmath}	
\usepackage{amssymb}	

\title[Noise reduction on transients via an autoencoder]{Noise reduction on single-shot images using an autoencoder}

\author[O. J. Bartlett et al.]{
Oliver. J. Bartlett,$^{1}$\thanks{E-mail: O.J.Bartlett-2018@hull.ac.uk}
David. M. Benoit,$^{1}$
Kevin. A. Pimbblet,$^{1}$
Brooke Simmons,$^{2}$
Laura Hunt$^{1}$
\\
$^{1}$E.A. Milne Centre, Faculty of Science and Engineering, The University of Hull, Cottingham Road, Hull HU6 7RX, UK\\
$^{2}$Department of Physics, Lancaster University, Bailrigg, Lancaster LA1 4YW, UK\\
}

\date{Accepted XXX. Received YYY; in original form ZZZ}

\pubyear{2021}

\begin{document}
\label{firstpage}
\pagerange{\pageref{firstpage}--\pageref{lastpage}}
\maketitle

\begin{abstract}
{We present an application of autoencoders to the problem of noise reduction in single-shot astronomical images and explore its suitability for upcoming large-scale surveys. Autoencoders are a machine learning model that summarises an input to identify its key features, then from this knowledge predicts a representation of a different input.
The broad aim of our autoencoder model is to retain morphological information (e.g., non-parametric morphological information) from the survey data whilst simultaneously reducing the noise contained in the image. We implement an autoencoder with convolutional and maxpooling layers. 
We test our implementation on images from the Panoramic Survey Telescope and Rapid Response System (Pan-STARRS) that contain varying levels of noise and report how successful our autoencoder is by considering Mean Squared Error (MSE), Structural Similarity Index (SSIM), the second-order moment of the brightest 20\% of the galaxy’s flux $M_2$$_0$, and the Gini Coefficient, whilst noting how the results vary between original images, stacked images, and noise reduced images. We show that we are able to reduce noise, over many different targets of observations, whilst retaining the galaxy's morphology, with metric evaluation on a target-by-target analysis. We establish that this process manages to achieve a positive result in a matter of minutes, and by only using one single-shot image compared to multiple survey images found in other noise reduction techniques.}

\end{abstract}


\section{Introduction}
Noise is omnipresent in astronomical imaging and beyond. Broadly defined, noise is the presence of unwanted artefacts within an image that can both litter and obscure the true target of interest contained within a given image. This may arise from both external physical sources such as incident cosmic rays affecting the image, or be internal to the physical hardware giving rise to the image itself, including the thermal properties of the hardware. For astronomical images, the presence of noise can have multiple effects including, but not limited to, downstream false segmentation, the incorrect computation of derived quantities such as (e.g.) galaxy morphology and brightness levels (fluxes, magnitudes, etc.), and the imposition of cosmic rays which may in and of themselves be reported as true (spurious) detections within the image. More generally, noise can cause blurred images and, in extremis, even completely block out the intended target of an image. This form of contamination is seen in most surveys, but is routinely observed in single-shot images (see, e.g., \cite{2015MNRAS.449..451W}). 

In some of our earlier work, \cite{CR_detection_and_noise_reduction} examined a variety of different ways in which noise could be identified and removed from single-shot images. That work investigated a variety of common, different algorithms to reduce the effects of cosmic rays. We summarize the four prime ones here.
Developed by \cite{Rhoads_CR_2000}, the first approach is to use an IRAF task that makes use of a linear filtering process that involves the image’s point spread function (PSF). This process applies a function to the image that is created from the difference between a Gaussian PSF and a scaled delta function. This identifies the cosmic ray pixels, since they are above a certain threshold. This algorithm iterates multiple times to ensure any and all cosmic ray pixels are identified. However it is worth noting that this process loses some precision when a cosmic ray is atop a target.
\cite{van_Dokkum_CR_2001} presents an alternative algorithm based on a Laplacian approach to cosmic ray identification (the L.A.Cosmic algorithm). This acts as an edge detection algorithm that involves parameters including the background standard deviation, rules for detecting neighbouring pixels of cosmic rays, and a discrimination between cosmic rays and target objects.
\cite{Pych_CR_2004}'s approach incorporates the use of histogram analysis on sections of an image. A frame is translated across the image, from which a histogram of the counts are interpreted. With a set threshold value, any cosmic ray pixels and neighbouring pixels are identified and removed from the original image. In turn this creates a cosmic ray map and cleaned image after the process has concluded.
Finally is the algorithm xzap developed by Dickinson (see within \cite{CR_detection_and_noise_reduction}). Their method applies a spatial medium filter that performs an unsharp masking process. This identifies pixels above a certain threshold. However this process also requires that the user manually inputs a section of the clear background sky and thus is dependent on input parameters that may not be accurate.
Thus there already exist a good number of algorithms to select between when deciding how to tackle noise reduction and cosmic ray rejection.

To determine how successful such approaches are, the nominal metrics of measurement for a comprehensive performance analysis include false detection rate, the quality of the reconstruction of the image, as well as the overall image denoising. 
The results from \cite{CR_detection_and_noise_reduction} suggest that the density of cosmic rays in an image does not impact the detection efficiency of said cosmic rays. Significantly, it is strongly linked by the density of real objects in the image. They find that the algorithm presented by \cite{Rhoads_CR_2000} produces the best performance when detecting cosmic rays, but the algorithm presented by \cite{van_Dokkum_CR_2001} has a high performance when it comes to both detecting and cleaning.

There are other methods of noise reduction that do not simply focus on cosmic ray removal for single-shot images. As noted in \cite{PS1_noise_reduction} amongst many other imaging works, stacking involves taking multiple exposures of the target and overlaying them. In doing so, it is possible to overlay each exposure to distinguish the target source and the noise source (see also \cite{article}) and hence remove the noise. Clearly though, for single-shot images or where transient events are of paramount importance, this is not an option.

\cite{2020A&A...643A..43R} further notes various noise removal techniques, from Gaussian smoothing, bilateral filtering [as introduced by \cite{Bilateral_filtering_1998} and adapted by \cite{Bilateral_filtering_2012}], total variation denoising, structure-texture image decomposition, and wavelet transforms. 

In this paper we specifically aim to address the loss in data from single-shot images, with application to retaining as much morphological image (i.e., about galaxies) as possible. The central question can be stated as: how do we reduce noise from images whilst retaining the morphological structure of the target? There are currently multiple ways in which noise can be reduced from an image, as illustrated above, and one newer approach is to involve machine learning. 

The practical applications of machine learning techniques for image processing of survey data are extensive. The IAC winter school 2018, documented by \cite{2019arXiv190407248B}, notes a plethora of topics from supervised and unsupervised machine learning, to evaluation methods on proposed models, and multiple supervised learning algorithms. \cite{2017arXiv170906257M} lists examples that can be used to illustrate the practicality of such techniques, such as showing the distinction between seven classes of variable stars using a confusion matrix.

Here, we make use of a known machine learning algorithm known as an autoencoder. Autoencoders have been used for many different applications in various different architectures. \cite{bank2021autoencoders} discusses the intricate details of many of these architectures, as well as general examples of how they have been applied. 
For instance, variational autoencoders have been used to generate new datasets by being trained on old ones. By making a link between labels and approximate representation of a target, autoencoders have also been attached to classification models.
\cite{mao2016imageRestorationAutoencoder} present their autoencoder model that reduces noise and can reconstruct images containing unwanted artefacts. This model is known as a denoising model, and is the focus of our work. 

These examples can be applied to the field of astrophysics in many different ways. \cite{2019MNRAS.485.2628L} introduces their convolutional denoising autoencoder (CDAE) to assist in the reduction of unwanted disturbances on the frequency dimension caused by overlooked foreground sources, as they attempt to uncover signals from the epoch of reionization (EoR) 

In this work, we will detail a new approach using a novel autoencoder to address the problem of noise in single-shot images. In Section \ref{section:Autoencoder}, we present a breakdown of relevant machine learning topics and concepts leading up to an explanation of autoencoders. Our methodology is laid out in Section \ref{section:Methodology and Data}, with Section \ref{section:Evaluation Metrics} going over how to effectively evaluate our results in a meaningful way, both visibly and numerically. Section \ref{section:Results} presents the outcome of running the model on our chosen dataset of galaxy images. Sections \ref{section:Discussion} and \ref{section:Conclusions} contain the discussion and conclusion of the work.

\section{Autoencoder}
\label{section:Autoencoder}

\begin{figure*}
\centering
\includegraphics[width=12cm]{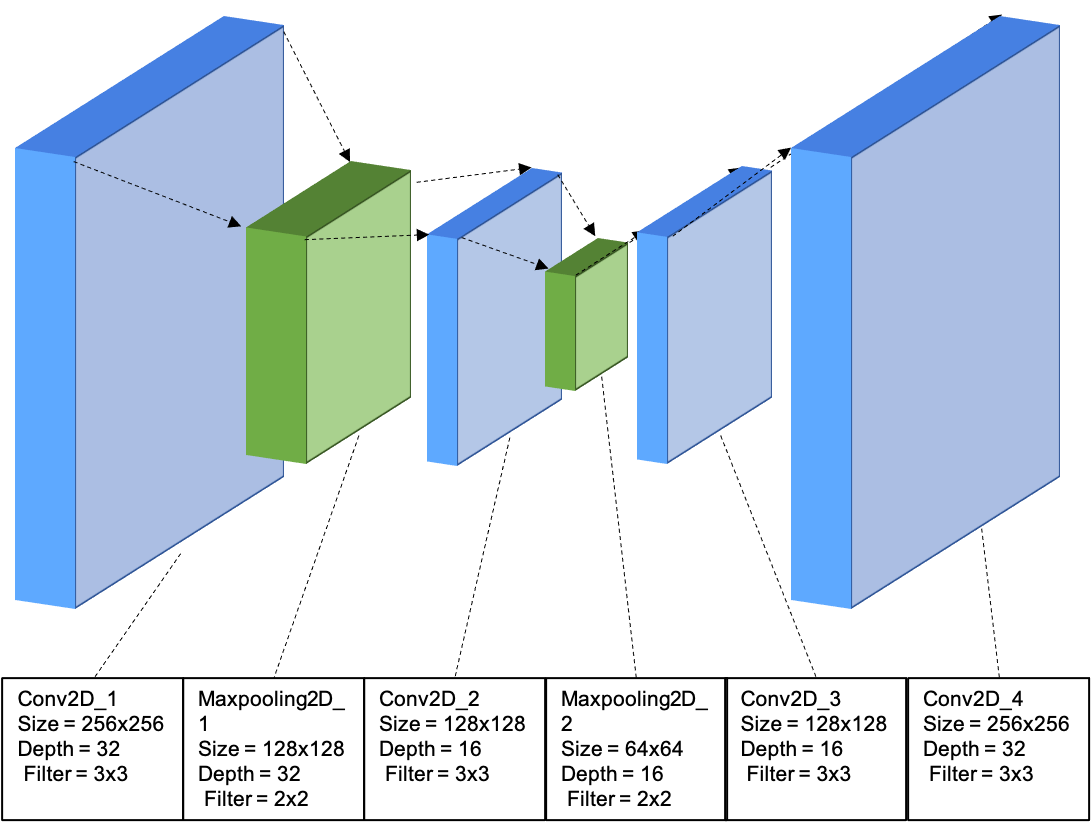}
\caption{A summarised version of the autoencoder model we have used. With our input layer on the left hand side, the autoencoder processing the assigned weights down to smaller and smaller layers, until it reaches the pinch layer in the centre. From this narrow neuron layer, the process is reversed, only using what it has attained thus far, and we reach the output layer on the right hand side.}
\label{own autoencoder model}
\end{figure*}

Simply put, an autoencoder is a specific formation of a neural network. In turn, a neural network is a model comprised of many connected layers containing neurons that conduct certain functions, as described by \cite{Montavon_2018}. 

Broadly, an input is presented to the network at the input layer, and depending on the settings of the network, certain processes are then done. In our case, a window (filter) is passed over areas of our image. At certain intervals the window conducts a calculation on the pixel values found within that frame (convolutional layers).
In each layer the neurons receive an activation from the previous layer, then more calculations and processes are conducted in those layers using the previous layer's results. This cascades down until the model has reached its end and an output is obtained. 

\subsection{Autoencoder model}

An autoencoder distinguishes key features of an input, encodes the input into a representation of said key features, and decodes the representation to produce an output that closely resembles the input, as listed by  \cite{2019MNRAS.485.2628L}. A key feature in our scenario would be objects with different pixel intensity values compared to their background and the noise mask applied. 

One of the main components of an autoencoder is that they possess a "pinch". In essence, the input progresses through the encoded layers of the autoencoder, in which the autoencoder reduces the input into a representation of the key features of the input. This representation is fully realised at the pinch, or bottleneck layer, as dubbed by \cite{bank2021autoencoders}. From this pinch layer, the representation is decoded to gain the output. The representation found in the pinch section does not display the features observationally, but rather as a pattern of pixel intensities. 

Once the autoencoder has this representation in the pinch layer, the entire processed is reversed. The output therefore is a un-summarised version of the representation found in the pinch layer, and it resembles the input as best as possible from this. This process is illustrated in Figure~\ref{own autoencoder model}, which shows our models structure and processes at different layers.

\begin{figure*}
\centering
\includegraphics[width=12cm]{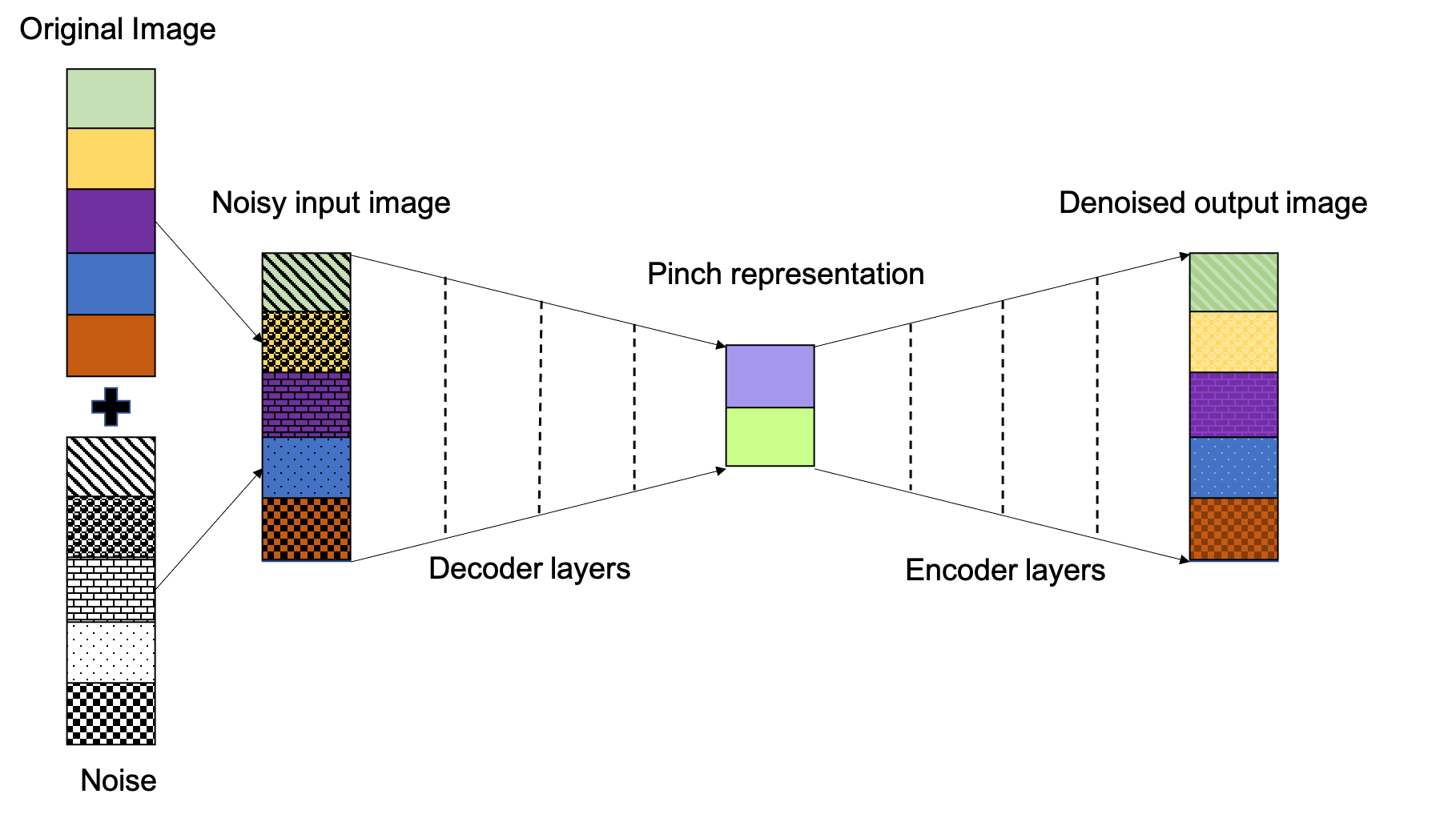}
\caption{A representation of the structure of a denoising autoencoder. this figure illustrates the effect of imparting noise onto an image, processing said image through the autoencoder, and presenting the outcome with a comparison to the original.}
\label{denoiser model structure}
\end{figure*}

With this model we train the autoencoder to identify key features of an image. This permits us to obtain a denoised image using predictions based on what the autoencoder has been trained on up until this point. By inputting noisy images we aim for the autoencoder to both recognise this noise and remove it as an output, as well as remove any  noise contained within the original image itself. This concept can be seen in figure \ref{denoiser model structure}, which is a simpler version of figure \ref{own autoencoder model}, but highlights the theory of our denoising autoencoder model.

\section{Methodology and Data}
\label{section:Methodology and Data}

In order to utilise an autoencoder to its maximum potential, it needs to be trained on a large and comprehensive dataset of single-shot images. We elect to use single images and via data augmentation create a dataset of said image at random orientations. 

The data that we elect to use for this purpose comes from the Panoramic Survey Telescope and Rapid Response System (Pan-STARRS1) (PS1) survey, explained in select papers, two of which are \citet{PS1_noise_reduction,2016arXiv161205560C}. Pan-STARRS is a wide-field imaging and data processing facility that examines and processes data from various astronomical topics, from solar system objects to the high redshift QSOs. The initial system used was labeled as Pan-STARRS1 (PS1), and the survey that this system conducted is what we will be using for our work. The reason for using PS1 survey data is that we sought a large suite of images that cover a wide area on the sky in order to cover a good range of different objects (of various intrinsic morphologies) across a variety of target densities that are representative of upcoming surveys, as well as an extensive range of images that cover the \emph{grizy} bands, and the kpc scale can be changed for each target so we can test the application of the autoencoder on different resolutions. For example, the Large Synoptic Survey Telescope (LSST) is due to be in operation in the near future, and will be able to probe 100 times more volume in the search for transient objects, as stated in \cite{LSSTsciencebook2-2009}, \cite{SKAsciencechallenge1-2018}, and by extension \cite{ASKAP-2008}, who discuss the Square Kilometer Array (SKA) and Australian Square Kilometer Array Pathfinder (ASKAP). All of these undertakings will provide an abundance of data and in turn data challenges, all due to the fact that it will be utilising the largest radio telescope in the world upon completion. 

\begin{figure}
	\includegraphics[width=\columnwidth]{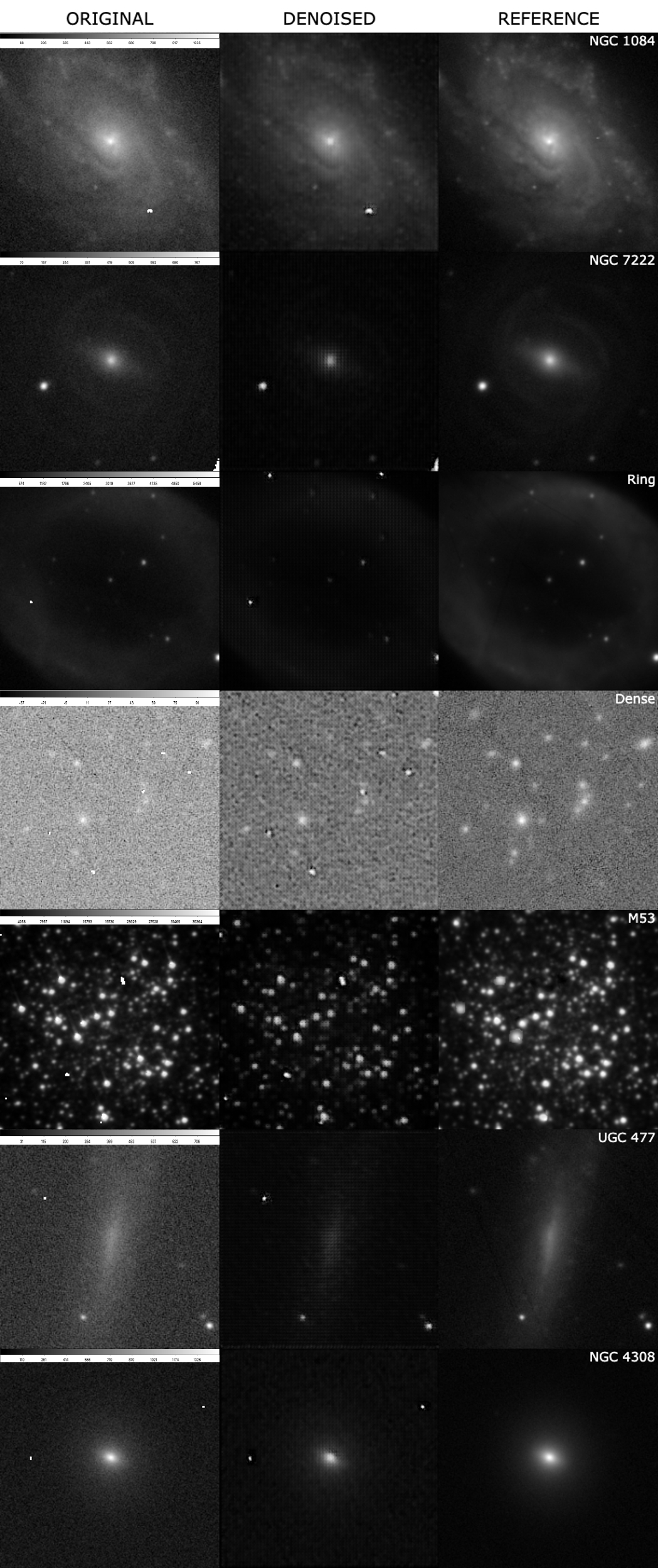}
    \caption{A portion of all of the images that have been applied to our denoising autoencoder process. They have been subjected to asinh scale adjustment for display purposes, and the flux scale in count/s is also displayed. The left hand column contains all of the single images we have taken from PanSTARRS. The central column shows the resultant denoised images. The right hand column shows the stacked images created in PanSTARRS. Going row by row down, we see NGC 1084, NGC 7222, the Ring Nebula, our dense cluster, M53, UGC 477 and NGC 4308. Also displayed are the flux scales for the Original images after they have been converted to 256x256.}
    \label{fig:Original denoised reference}
\end{figure}

The images that we have chosen from PS1 can be seen in the left hand column of figure \ref{fig:Original denoised reference}. These images can be found via the PS1 Image Cutout Server. In brief they represent a broad range of target morphologies, consisting of a spiral galaxy, a barred spiral galaxy, the Ring Nebula, a cluster of galaxies, Messier 53 -- a globular cluster, a low surface brightness galaxy, and an elliptical galaxy. We have also included targets of similar nature to NGC 7222, The Ring Nebula, Dense, and M53 to test if the autoencoder is type dependent. These targets are NGC 64. NGC 6543, The Virgo Cluster, and NGC 288. Our input images are 240x240 FITS image cutouts, and the band and exposure time for each image can be seen in table \ref{table:Image data}.

Our choice of images depended on what was available on a case-by-case basis. For instance, certain bands for certain targets did not have a full picture of said target; these images would contain gaps of missing pixels due to detector issues or non-useful pixel areas. Therefore with this in mind we chose images that contained a complete map of the target on a random basis for bands. The exposure times for each warp can be found in table \ref{table:Image data}, as well as in the header of the fits file itself. 

A full list of the data can be found in table \ref{table:Image data}. For each target, we obtain not only a final stacked image, but also an individual observation frame from within the stack. The stacked images are created and stored in this server alongside the individual images that were used to create said stack. Both of these are displayed in figure \ref{fig:Original denoised reference}. The former will ultimately give us an indicator of the resultant image that we want to obtain from the latter, with the strong caveat that it is impossible to recover the stacked image from an individual frame. 

\begin{table*}
	\centering
	\caption{The band and identification number of each of the single-shot images we have taken from the PS1 image access. The stacked versions of the images used belong to the same band.}
	\label{table:Image data}
	\begin{tabular}{lcccccr} 
		\hline
		  Target & Band & ID & RA (Deg) & Dec (Deg) & Exposure Time (s)\\
		\hline
		 NGC 1084 & y & 1153.012 55786.62506 & 41.450 & -7.578 & 30\\
		 NGC 7222 & g & 1405.053 55799.35382 & 332.716 & 2.106 & 43\\
		 Ring & y & 2069.026 55460.27368 & 283.396 & 33.029 & 30\\
		 Dense & i & 1805.032 56817.27208 & 176.662 & 21.281 & 45\\
		 M53 & i & 1725.051 56355.61523 & 198.230 & 18.169 & 45\\
		 UGC 477 & i & 1681.087 55457.37481 & 11.5547 & 19.490 & 45\\
		 NGC 4308 & y & 1971.057 55214.62499 & 185.487 & 30.074 & 30\\
          NGC 288 & z & 717.034 55566.20286 & 13.198 & -26.590 & 30\\
          NGC 64 & g & 1144.024 56214.34293 & 4.377 & -6.825 & 43\\
          NGC 6543 & y & 2528.063 55728.47759 & 269.639 & 66.633 & 30\\
          Virgo Cluster & i & 1636.006 56002.56836 & 187.697 & 12.337 & 45\\
		\hline
	\end{tabular}
\end{table*}

A key point to recall here is the nature of the single-shot images we wish to reduce noise from: for a single target during a single-shot survey, there will not be a large enough dataset available to train and test an autoencoder on. 
Therefore data augmentation must be incorporated. In other words, we take the input image -- one of the individual frames noted above -- and from this spawn many further images to train with. To achieve this end, we take the images and randomly rotate them through arbitrary angles and apply modest translations to them.

The single-frame input images may also contain noisy, transient objects, as would be expected from surveys such as PS1 and beyond. We emphasize that if we input a noisy single-shot image into our autoencoder, our resultant output should still be similar to the input, i.e., the original noisy single-shot image. 

In order to output a cleaner image than our input, we incorporate and adapt the concept of a denoising autoencoder.
We divide our data up in to training and testing, specifically a 70/30 split. These sets are comprised of our original, noisy, single-shot images taken from stacks, as well as a noisier version of these images that we have deliberately added extra (statistically known) noise to.
We have applied random Gaussian noise, Poisson noise, and read noise, in order to simulate the conditions found in single-shot images. This has been done in other denoising projects. \cite{U-net-denoise-2020} incorporate photon shot noise, dark noise, and read-out noise to create their short exposure images, in which they use U-nets to denoise said images. \cite{Different_denoising_methods_using_Gaussian} apply Gaussian noise to simulated survey images, which they then run through different forms of denoising algorithms to test each methods ability to reduce noise, detect the target source, and preserve the pixel intensity and morphology of said target. \cite{CR_detection_and_noise_reduction} present the results of different noise reduction models being applied to Image Reduction and Analysis Facility (IRAF) images, as well as real data. For their simulated images they focus on introducing cosmic rays, but alongside that they include Poisson noise and read noise set at 5 electrons. \cite{PS1_noise_reduction} details the detections and image processes performed by the Pan-STARRS1 Science Consortium, in which they discuss the effects of gaussian, read and Poisson noise. 
For these reasons presented we chose to follow suit and incorporate simulated read, random Gaussian, and Poisson noise into our methodology. This covers noise that appears on an image due to the transformation of signal detections to a digital format, as well as the impact of temperatures, electronic issues, and the fluctuations of photons impacting the detector. 
For random Gaussian noise we have set the noise intensity at 10\%. 
The Poisson noise is created from the data itself, as well as using a module called \verb'skimage.util'. We have set the read noise level to 5 electrons of gain 2. 
We use a numpy package called \verb'random.normal' to simulate read noise. It uses the probability density function of the Gaussian distribution. For inputs are scale, shape and gain. Scale indicates the normal distributions standard deviation, which is calculated by the division of the amount of read noise in elections and the gain of the camera, in units of electrons/ADU. Size simply dictates the output shape, which in our case we want it to be the same as its input. 

We train our autoencoder using these two sets, each set amassing to a total of 3000 images. From these sets we ask the autoencoder to predict the output if we specify what its input would be. 
Having been trained on datasets that comprise of normal and extra noisy data, this results in the autoencoder identifying noise, and therefore its prediction should in principle contain less noise than the input. 

Our hypothesis is that the autoencoder will recognise noise as an unwanted feature whilst preserving target features (e.g., galaxies). Hence it should reduce not only the random Gaussian noise, but the original noise as well, thus leaving us with an output that has an improved signal-to-noise ratio than that or its original counterpart. 

The final issue to resolve is the well known neural network problem of overfitting. Outlined by \cite{overfitting}, machine learning models develop complex patterns between input and output data. However, if there is a limited amount of data to produce these patterns, these patterns only exist in the finite space between the used data, and can not be applied to other datasets, even if taken from the same source. A way to reduce the effects of overfitting in our scenario is to randomise the training and testing sets. This way the model does not attribute rotation or translation as an identifiable pattern and risk overfitting. This method can be seen in \cite{DataShuffleOverfitting2019}. Through our use of data augmentation on our images, this also assist in overcoming the overfitting issue. We note that we forgo the use of dropout layers in our model, which are designed to combat overfitting, as this is counterproductive to the steps we have already taken with our data and model. 

Our model is adapted from a keras example, a link to which can be found in the Data Availability section at the end of the paper. As seen in figure \ref{own autoencoder model}, our model contains convolutional and maxpooling layers in the decoder and encoder sections of the autoencoder, with transpose versions in the encoder section to reverse the process of the decoder. As stated by \cite{Convolutional_and_max_pooling_layers_2015} the role of a convolutional layer is to identify local conjunctions of features from the previous layer in the network, and the role of a pooling layer is to combine homogeneous features into one. \cite{ConvolutionalNN_2018} goes into more detail on the nature of convolutional and max pooling layers. They describe a convolutional layer as a key component for feature extraction, and the layer itself is comprised of linear and non linear operations. The linear operation in this case would be the convolution operation, the first step of which is creating a feature map of the input. This is done by using different kernels with the tensor applied across the input. The result is a feature map that is an array smaller than that of the input. A parameter that defines a convolutional operation is the size and number of kernels. For ours we use a simple $3\times 3$ matrix that is optimised by the neural network. This helps describe the linear convolution operation, and once set and passed through, the outputs progress through the nonlinear section. This nonlinear operation involves the activation function. For this we used the rectified linear unit (ReLU - detailed and used by \citealt{ReLU2015} and \citealt{ReLU2019}). In order to reduce the dimensions of the input effectively as it travels through the network we incorporated certain pooling layers. In our work we used Max pooling layers, specifically \verb'Maxpooling2D' (\citealt{MaxPooling2020}, \citealt{MaxPooling2012} (originally conceived of in \citealt{MaxPooling1999}). Max pooling keeps the maximum value of random patches from the feature map, and discards the rest. Our max pooling layers have a window size of $2\times 2$. We have also incorporated Adam (\citealt{AdamOptimiser2017}) and binary\_crossentropy (\citealt{BinaryCrossEntropy2020}) for our optimizer and loss functions. 

Described in its original paper by \citealt{Adam_optimizer_Kingma_2015}, the optimizer is simple to put into the network, requires little memory to be effective and efficient, good in cases of large datasets, and handles noisy gradients well.
provided by the Keras models package, Binary crossentropy calculates the cross entropy loss for a binary task. Since we normalise our target images to be between 0 and 1 so that they can be inputted into the network, we consider intensities as varying ways of “yes” and “no”. In a way we approximate it as if it is a binary decision map, where we assume the background is 0, target is 1, and everything else is in between. 

We decided to incorporate these as they are the most common applications used to construct a neural network. We acknowledge that whilst Mean Squared Error may have been better suited to construct our model around due to its use as a metric, this is an initial proof of concept piece to highlight that it can do well even with the basic features.

\renewcommand{\thefigure}{\arabic{figure} (Cont.)}
\addtocounter{figure}{-1}
\begin{figure}
	\includegraphics[width=\columnwidth]{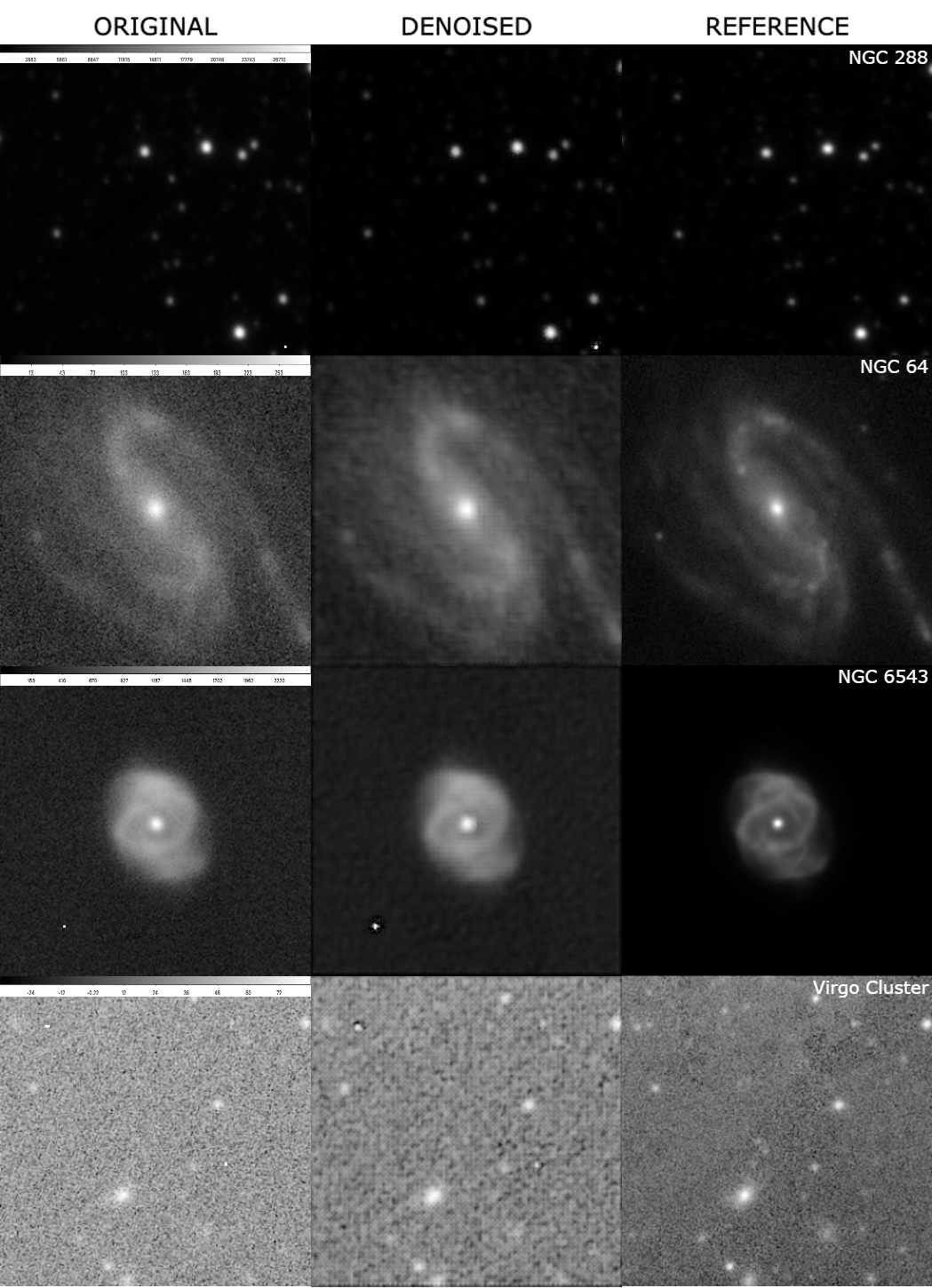}
    \caption{A portion of all of the images that have been applied to our denoising autoencoder process. They have been subjected to asinh scale adjustment for display purposes, and the flux scale in count/s is also displayed. The left hand column contains all of the single images we have taken from PanSTARRS. The central column shows the resultant denoised images. The right hand column shows the stacked images created in PanSTARRS. Going row by row down, we see NGC 288, NGC 64 NGC 6543 and the Virgo Cluster. Also displayed are the flux scales for the Original images after they have been converted to 256x256.}
    \label{fig:Original denoised reference new targets}
\end{figure}
\renewcommand{\thefigure}{\arabic{figure}}

\section{Evaluation Metrics}
\label{section:Evaluation Metrics}
In order to define the effectiveness of our noise reduction autoencoder we must select our metrics of merit. If we have data that has had its noise reduced by other means, we could visually compare our outcome with this. However, this is solely based on the human eye and what we would perceive as a difference, which is different from person to person. Furthermore, there may be segments that have been removed surrounding a galaxy that may actually belong to the morphology of said galaxy. Thus we need metrics that can statistically identify change in noise contamination and simultaneously consider that less noise is a good change if the targets morphology remains intact. Accordingly, we  consider and incorporate a variety of metrics that include: (a) the Gini coefficient; (b) the second-order moment of the brightest 20\% of the galaxy’s flux $M_2$$_0$; (c) Mean Square Error (MSE); and (d) Structural Similarity Index (SSIM). 

First seen in \cite{Gini_firstused_1912}, and demonstrated by \cite{GINI2003}, the Gini coefficient is a useful tool that can describe the overall morphology of all galaxy types. The Gini coefficient measures the distribution flux on a pixel level throughout the image it is analysing, as seen in the works of \cite{GINI2018}. This can be seen as similar to the concentration metric used in CAS \cite{Abraham_CA_firstused_1994}, but as noted in \cite{GINI2021} the Gini coefficient is more sensitive to the small changes in pixel values. This is again highlighted by \cite{GINI2003} where they measure the Gini coefficient of local galaxies found in SDSS, and state that the Gini coefficient is very similar to the measurement of the targets' central concentration. The difference between them is that Gini identifies a significant scatter, which is due to changes in the mean surface brightness of galaxies. First introduced in \cite{Gini_firstused_1912}, later adapted into the field of observations by \cite{GINI2003}, and further described by \cite{GINI2018}, the equation used to calculate the Gini coefficient, stating that a value of G=1 means all the light in an image is concentrated in one pixel, whilst a value of G=0 means that the light has been equally distributed across the entire image. the formula they provide is referenced in equation \ref{gini Equation}. We note that the Gini coefficient code we have used is different to the equation we have listed here. The equation the code is based off of can be found in the reading material listed in section \ref{section:Data availability}. There the author states that their equation is an estimate, and that by multiplying their equation by $n/(n-1)$ this gives the results expected from the equation listed here in this paper. However, due to the size of our images, the conversion results are negligible. 

\begin{equation} \label{gini Equation}
G = \frac{1}{\hat{I}n(n-1)}\sum_{i}^{n} (2i - n - 1) \hat{I_i}
\end{equation}

where \^{I} is equal to the mean flux of the image pixels, $\hat{I}_{i}$ is the flux in the $i^\textrm{th}$ pixel, and $n$ is the total number of pixels in said image, as explained in \cite{GINIequation2015} as well. 

SSIM and MSE are intrinsic to the autoencoders model, and therefore we evaluate these metrics on the full image. In the cases of Gini and $M_2$$_0$ however we apply a segmentation map to retain the target and exclude the background. The segmentation map removes all non target pixels, meaning all measurements are solely conducted upon the targets. We have applied this process to a number of our targets, but we do note that value for sigma in the code is altered from 1.5 to 0.0 for Ring, Dense, M53, NGC 288 and the Virgo Cluster. This is due to the fact that the segmentation map is not designed to isolate clusters of targets, and altering this value produced a result that encompassed as much of the key features of the targets as possible. For the segmentation process we kept the centre of the galaxy's coordinates to be 128x128, and the map was created on the Original image and then applied to the other images. We found for most targets this position was approximate enough to its centre, and in the case of clusters we decided to leave the centre value unchanged. 

Described by \cite{Gini_M20_Lotz_2004}, $M_2$$_0$ is the second-order moment of the brightest 20\% of the target. 
To calculate $M_2$$_0$ we first need to calculate $M_t$$_o$$_t$, described in equation \ref{Mtot Equation}.  

\begin{equation} \label{Mtot Equation}
M_{tot} = \sum_{i}^{n} M_{i} = \sum_{i}^{n} f_{i} [(x_{i}-x_{c})^2 + (y_{i}-y_{c})^2]
\end{equation}

where x$_i$ and x$_c$ represent the pixel coordinates of the selected pixel and the pixel designated as the image's chosen centre, all in the x axis. This is the same for y$_i$ and y$_c$. f$_i$ is equal to the selected pixel's intensity. 

Once we have calculated $M_t$$_o$$_t$ we can calculate $M_2$$_0$ via the equation \ref{M20 Equation} below.

\begin{equation} \label{M20 Equation}
M_{20} \equiv \log _{10} \left(\frac{\sum_{i} M_{i}}{M_{tot}}\right) , while \sum_{i} f_{i} < 0.2 f_{tot}.
\end{equation}

As we did with Gini, we apply the $M_2$$_0$ metric to targets that have undergone the segmentation map process.

Our final pair of metrics are the Mean Squared Error (MSE) metric, and the Structural Similarity Index (SSIM). Unlike the above, these are not morphological measurements in and of themselves but a measure of the similarity of two images.

The MSE metric finds the difference between the images by subtracting the pixel intensities, squaring these differences, summing these and dividing the sum by the total number of pixels. A mean squared error value of zero means that both images in a pair are perfectly similar: there is no difference in the intensity of pixels between these images. The equation for this metric is seen in equation \ref{MSE Equation}, below.

\begin{equation} \label{MSE Equation}
MSE = \frac{1}{MN} \sum_{x=1}^{M} \sum_{y=1}^{N} [u_{xy} - v_{xy}]^2
\end{equation}

As stated by \cite{MSEequation2017}, $M$ and $N$ represent sizes of the two images that are being compared, with $u$ and $v$ representing said images. $x$ and $y$ here indicated the coordinates of the two images. 

SSIM views the images as a whole via the use of windows placed over the image and takes in the average balance of pixels per window. The difference between the two images via the windows formulates the measure as a statistical measure of similarity. These windows are of size 11x11 pixels, as stated by the documentation for \emph{skimage.metrics}. Equation \ref{SSIM Equation} describes how to calculate SSIM. 

\begin{equation} \label{SSIM Equation}
SSIM(x,y) = \frac{(2 \mu_x \mu_y + c_1)(2 \sigma_{xy} + c_2)}{(\mu_x^2 + \mu_y^2 +  c_1)(\sigma_x^2 + \sigma_y^2 + c_2)}
\end{equation}

Found in the works of \cite{SSIMequation2016} and \cite{SSIMequation2019}, $c_1$ and $c_2$ represent constants, $x$ and $y$ represent the two images that are being compared. $\mu_x$ is the average of $x$, and this is the same for $\mu_y$ and $y$. This is also the case with $\sigma_x^2$, with this being the variance of $\mu_x$. Noted in \cite{SSIM_image_quality_assesment} and \cite{SSIM_love_or_hate} the constants are for the purpose of stabilising the equation in the situation where certain values approach zero. These constants are derived from the range of pixel values used, as well as another constant that is less than or equal to one. Finally  $\sigma_{xy}$ is  the covariance of $x$ and $y$. For the results of the metric, if the value of the compared image is closer to 1, the images are more similar. If they are closer to 0 they are more different. 

\section{Results}
\label{section:Results}

Here we present the visual and statistical results of our work. From here we have our original image (the single-shot image taken from a stack found in PS1), the denoised image that is the resultant of the autoencoders processing of the noisy image, which is the original image with artificial noise placed upon it, and our reference image (the stack image taken from PS1).

Firstly we observationally examine the results of our own work before referring to the metrics used. Nearly all of the images can be seen to have an improvement in noise reduction and target morphology structure. Whilst we knew noise could not be completely removed from a single-shot image, the reduction of noise across all of the images is evident. Prime examples of this can be seen with NGC 1084, M53, UGC 477, NGC 4308, NGC 64. NGC 6543 and the Virgo Cluster. All of the other targets also exhibit this, but not as prominently. One worth noting is our dense cluster image can still be seen with some amount of noise, but it has done well in highlighting the targets scattered throughout the image. We believe that it did not do as well with noise reduction as a result of the pixel intensity of the targets being too low, and the noise atop them being too prominent, that the autoencoder struggled somewhat telling them apart. 

In terms of MSE we refer to table \ref{table:MSE}, where we have to bear in mind two things; that the closer the value is to zero, the more alike the images are in terms of MSE, and that even if the difference in values are small, this still resembles a change. For NGC 1084 we see that the lowest value belongs to the comparisons between referenced and Denoised, implying that our denoised image resembles the stacked image more so than the original image. This means that our denoised method has done its job correctly in terms of MSE and reduced noise down significantly compared to its original state, whilst retaining the targets morphology. This scenario can also be seen for NGC 7222, Dense, UGC 477 and NGC 4308. M53's denoised image is not as similar to our reference image than our noisy or original image is, and this may be due to the multiple targets that make up the image, or that our denoised M53 image has a noticeable difference in pixel intensity compared to its original and noisy counterpart. Our denoised Ring image is not quite as close to the reference image than our original image, and this could be again due to pixel intensity differences in each image. The Original vs Denoised, Reference vs Denoised and Reference vs Original results for NGC 288 are all very close. There are less point sources compared to similar targets we have observed, such as M53, and this could be why the difference between results is so small. For the case of NGC 64 our Denoised image resembles the Original image more so than the Reference image. The Reference image however has less pixel intensity in the ring and bar of the galaxy compared to what we can see in the Original and Denoised image. Therefore this could be a reason why our Denoised image is more similar to our Original image. This is the same for NGC 6543. For the Virgo Cluster we can see our Denoised image is more like the Reference image opposed to comparing the Original image to the Reference image. Implying noise has been reduced overall. 

\begin{table*}
	\centering
	\caption{The MSE results. The closer the values are to 0, the more alike the compared images are to each other. For a positive result we expect our denoised image to be most similar to the reference, or at the very least very different from the noisy image.}
	\label{table:MSE}
	\begin{tabular}{lccccccccccr}
		\hline
		   & NGC1084 & NGC7222 & Ring & Dense & M53 & UGC477 & NGC4308 & NGC288 & NGC64 & NGC6543 & Virgo Cluster\\
		\hline
		 Original vs Noisy & 0.0027 & 0.0034 & 0.0033 & 0.0043 & 0.0028 & 0.0034 & 0.0030 & 0.0046 & 0.0029 & 0.0027 & 0.0034\\
		 Original vs Denoised & 0.0014 & 0.0009 & 0.0003 & 0.0105 & 0.0022 & 0.0045 & 0.0005 & 0.0000 & 0.0003 & 0.0003 & 0.0041\\
		 Reference vs Noisy & 0.0048 & 0.0051 & 0.0035 & 0.0244 & 0.0033 & 0.0093 & 0.0050 & 0.0050 & 0.0091 & 0.0088 & 0.0261\\
		 Reference vs Denoised & 0.0004 & 0.0006 & 0.0004 & 0.0093 & 0.0035 & 0.0005 & 0.0003 & 0.0001 & 0.0045 & 0.0025 & 0.0197\\
		 Reference vs Original  & 0.0010 & 0.0010 & 0.0003 & 0.0323 & 0.0015 & 0.0029 & 0.0006 & 0.0001 & 0.0037 & 0.0038 & 0.0252\\
		\hline
	\end{tabular}
\end{table*}

Results for the SSIM metric are shown in table \ref{table:SSIM}. Similar to when we discussed the MSE results, we restate that if the resultant value is closer to one, the more similar the images are to each other for this metric. Our denoised results are most similar to the reference images in the cases of NGC 1084, Dense, UGC 477, NGC 4308, and NGC 288. Therefore the noise reduction on these targets when using SSIM as a metric of comparison has been a success. NGC 7222 has our denoised image and our original image being very similar to the reference, with the original being slightly more similar. This is also seen somewhat in for Ring and M53. A potential reason for this may be again due to the intensities in the denoised images. It may be also due to the size of the windows used, as these cases contain the least and objects in their images. NGC 64, NGC 6543 and the Virgo Cluster show that the Denoised image is more similar to the Reference image compared to the Original image. But the Denoised image is more similar to the Original image. We still see this in a positive light in that with the reference as a benchmark the Denoised image has improved over the Original image. 

\begin{table*}
	\centering
	\caption{The SSIM results. Comparing images under the SSIM metric produces these results, where if the value is closer to 1 the more similar the images are to each other. Having a denoised image be more similar to the reference image, and/or for the denoise image to be very different from the noisy image is evident our model is functioning correctly.}
	\label{table:SSIM}
	\begin{tabular}{lccccccccccr} 
		\hline
		   & NGC1084 & NGC7222 & Ring & Dense & M53 & UGC477 & NGC4308 & NGC288 & NGC64 & NGC6543 & Virgo Cluster\\
		\hline
		 Original vs Noisy & 0.62    & 0.51 & 0.48 & 0.77 & 0.55 & 0.59 & 0.55 & 0.19 & 0.62 & 0.61 & 0.78\\
		 Original vs Denoised & 0.90    & 0.70 & 0.86 & 0.59 & 0.85 & 0.36 & 0.84 & 0.98 & 0.93 & 0.97 & 0.62\\
		 Reference vs Noisy & 0.50    & 0.36 & 0.48 & 0.37  & 0.64 & 0.26 & 0.29 & 0.14 & 0.34 & 0.11 & 0.36\\
		 Reference vs Denoised & 0.97    & 0.89 & 0.85 & 0.65  & 0.71 & 0.88 & 0.96 & 1.00 & 0.65 & 0.35 & 0.60\\
		 Reference vs Original & 0.89    & 0.91 & 0.99 & 0.44  & 0.94 & 0.56 & 0.75 & 0.99 & 0.64 & 0.28 & 0.47\\
		\hline
	\end{tabular}
\end{table*}

\begin{figure*}
	\includegraphics[width=15cm]{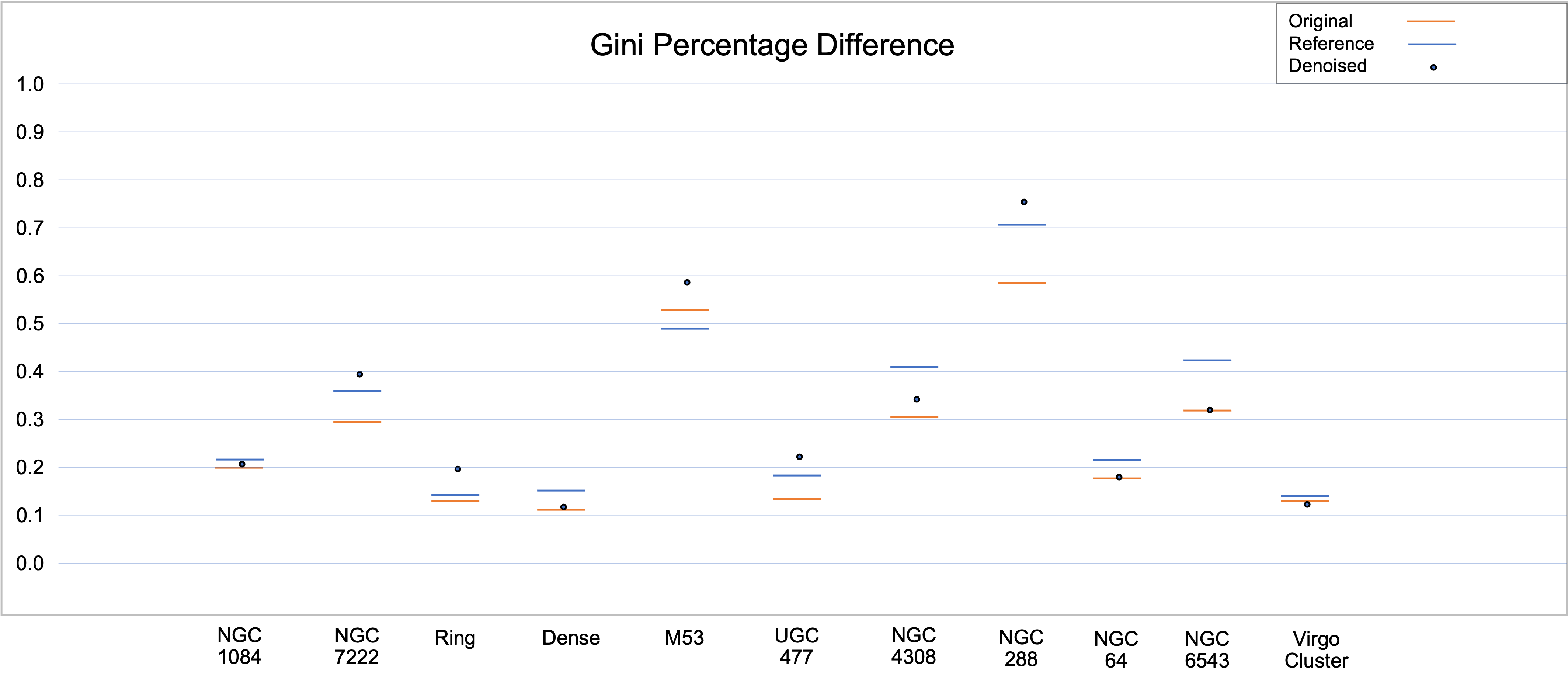}
    \caption{The Gini results for all targets after the segmentation map process; the Original image, denoised image and reference image. Due to each target being significantly different in morphology and intensity, we focus more on the relationship of each targets image, rather than a comparison of targets as a whole.}
    \label{fig:Gini segmentation map percentage metric}
\end{figure*}

With the Gini coefficient we recall that G=0 equates to each pixel in the image containing the same level of intensity, and G=1 equating to a single pixel containing all of the intensity in said image. This metric is slightly more in depth as we evaluate each image as a separate case due to its morphology and intensity; on average low intensity galaxies should have a Gini coefficient closer to 0, but with noise this may be higher or lower depending on the form of noise found. In our case our noise is spread throughout the image, so in most situations the noise should lower our Gini values. For the case of potentially increasing the Gini value, cosmic rays would be a prime example. In an attempt to combat this we only apply Gini to the versions of the targets that have undergone the segmentation map process. We therefore briefly examine each case individually. These results can be cross examined with figure \ref{fig:Gini segmentation map percentage metric} and table \ref{table:Gini segmentation}. The plot shows how the denoised image of a target tends to resemble either the original or reference image. 

\begin{itemize}
  \item NGC 1084 - The Denoised Gini value lies in between the Original Gini value and the Reference Gini value. The separation between the values is near even. However the Denoised value is slightly closer to the Reference value. This is a positive trend; it is more similar to our Reference image than our Original image. 
  \item NGC 7222 - Here we see that the Denoised target has the highest Gini value in this case. However it is closer to the Reference value than the Original value. Therefore, whilst not in between the Original and Reference value, the fact that the Denoised value is closest to the Reference value is still seen positively. 
  \item Ring Nebula - The Gini value of our denoised Ring Nebula again is closest to the Reference value than the Original. This implies that, in terms of this metric, the image that has been denoised from the autoencoder process is more similar to the Reference image produced by stacking, compared to the Original single-shot image contaminated with noise.
  \item Dense cluster - The denoised Gini value is one of the lowest. The reason for this is unclear. Perhaps the amount of sources with varying intensities confused the autoencoders performance with light concentration. That taken with the fact that the segmentation map highlighted multiple point source tells us that the denser the target image, the more the autoencder may struggle.
  \item Messier 53 - The denoised value is one of the highest here, implying that the noise around the target point sources have been reduced, making each point source contain more pixel intensity over the entire image. However it is also closer to the Original Gini value than the Reference Value. Coupled with our discussion on the Dense cluster, future investigations and considerations for this type of target should be implemented. 
  \item UGC 477 - The Gini value lies higher than the Original and Reference value. However, like NGC 7222 and the Ring Nebula, the Denoised value is closer to the Reference value. Once again, we take this as a positive result. 
  \item NGC 4308 - This is a similar case to NGC 1084, however the seperation between the Original and Reference values are greater. We also see that the Denoised value favours slightly more towards the Original value. With that said, it still lies between the two other values, and therefore this is a positive result.
  \item NGC 288 - The Denoised value is above, but closest to, the Reference value, and there is a large separation between them and the Original value, with the Original value being much lower. This would imply that noise has been reduced on the entire target and the point sources are clearer since they would contain the majority of pixel intensities across the target image. 
  \item NGC 64 - We see that the Denosied value lies in between the Reference and the Original value. Whilst favouring the Original, this is still a positive result. This however is evidence to back our claim that our autoencoder model can reduce noise significantly on a single image, but will not be able to fully achieve the clarity that stacking can provide most of the time. 
  \item NGC 6543 - Similar to previous targets, the Denoised value lies in between the Original and Reference values, but greatly favours the Original value. We therefore see this as a positive result overall.
  \item Virgo Cluster - The only Denoised Gini value that is below the Original and Reference values. Seen in a similar manner to our Dense target, wherein multiple targets with a considerable amount of noise already on the image results in the autoencoder being overwhelmed. 
\end{itemize}

The results from the Gini metric show that the model works as intended; to reduce noise across the image. A point for future experimentation is the effect it has on the intensity of the target pixels. In certain cases this is not seen, and others it is. This may be due to the morphology of the target. Narrowing down what targets this is seen on would be an interesting step forward in the future, as it seems certain clusters are prone to this, whereas ellipticals and spirals are not. The results of Gini when applied to the segmentation maps of the targets reinforce this point, as well as the $M_2$$_0$ results demonstrate a positive trend. 

 As the $M_2$$_0$ metric allows us to evaluate how intensity of a target is distributed whilst considering the target's total pixel intensity spread for a certain total percentage, we conduct our $M_2$$_0$ evaluation approach to that of our Gini evaluation; a positive result involves the Denoised value being in between the Original and Reference value, and/or closer to the Reference value. we will once again discuss each target individually. Results for $M_2$$_0$ can be seen in table \ref{table:M20 segmentation}. We recognise that this process is not designed for certain targets, such as clusters, and have altered the settings of the code to account for this as much as we could. 

 \begin{itemize}
     \item NGC 1084 - The Denoised value lies  between the Original and Reference values. Better still it is closer to the Reference value. This is an ideal result, and therefore a success for this target when examined by this metric.
     \item NGC 7222 - Here we see that the Denoised value has exceeded past the Original and Reference value. However it is closest to the Reference value, and we can take this as a positive result. This follows its Gini counterpart in a similar trend, and this could mean that for this target the pixel intensity is perserved for the morphology of the target in the segmentation map, but due to the map being taken from the Original image this provides different features to be highlighted and considered.  \item Ring Nebula - We see that the Denoised value lies between the Original and Reference value. And we can see that it is closest to the Refrence value, and therefore meets the criteria for the ideal case of denoising. 
     \item Dense Cluster - Similar to some previous targets, the Denoised value lies between the Original and Reference values. However in this case it is closer by a significant amount towards the Original value. This is still a positive result based on its position still.
     \item M53 - The Denoised value is is closest to the Original value, and not in between the Original value and Reference value. The difference between all values is small compared to other targets however. M53's segmentation map has also been altered to capture as much of the point sources as possible. Therefore whilst we acknowledge the result, we also state that this examination is not ideal for this target type. Investigations into autoencoder application for clusters would be needed in a future paper, as well as potential other metrics specific to this type of target. 
     \item UGC 477 - Here we see that the Denoised value is not in between the Original and Reference value, but it is closest to the Reference value. We do note the larger difference between the Denoised and Reference value and the Original and Reference value, even though we also consider this to be an overall positive result.
     \item NGC 4308 - We see a positive position for the Denoised value, wherein it is extremely close to the Reference value. Even though it is not in between the Original and Reference value, it is so close to the Reference that we can see this in a positive position. 
     \item NGC 288 - The Denoised value is the lowest for this target, but closest to the Reference value. We can see that the Original and Reference values have a smaller separation compared to the Denoised and Reference value, but compared to other targets these differences are minuscule. 
     \item NGC 64 - In a similar position to its comparative partner, NGC 7222, the Denoised value is closest to the Reference value, but is separated greatly from said Reference value. Again, due to its comparative partner, we suggest future papers to investigate the denoising affect on pixel intensities for such targets. 
     \item NGC 6543 - Here the Original and Denoised values are extremely close together. As well as that, the Denoised value is not in between the Original and Reference value. The Reference image is drastically different to the Original, and as much as the autoencoder has removed noise, it can not achieve the same results as the stacking process with only one image at its disposal. However we reiterate its abilities to denoise targets to such an extent with only one image available to it, whilst perseving the targets pixel intensity and morphology. 
     \item Virgo Cluster - The Denoised value is in between the Orginal and Reference values, but is closer to the Original value. Therefore even for this cluster, we see a positive result. 
     
 \end{itemize}

\begin{table*}
	\centering
	\caption{The Gini results when calculated on the segmentation map of the original target. This means only the target is considered, not background or non-target objects. Each target is discussed individually, but again we aim for our denoised image to be as close to the reference image as possible. Values approaching 0 means the intensity of the pixels in the image are more spread out over the whole image, and values approaching 1 means the intensity of the pixels in the image is all contained in one pixel.}
	\label{table:Gini segmentation}
	\begin{tabular}{lccccccccccr} 
		\hline
		   & NGC1084 & NGC7222 & Ring & Dense & M53 & UGC477 & NGC4308 & NGC288 & NGC64 & NGC6543 & Virgo Cluster\\
		\hline
		 Original & 0.201 & 0.296 & 0.134 & 0.117 & 0.531 & 0.135 & 0.307 & 0.586 & 0.178 & 0.323 & 0.132\\
		 Denoised & 0.213 & 0.391 & 0.193 & 0.124 & 0.587 & 0.225 & 0.341 & 0.755 & 0.185 & 0.326 & 0.122\\
		 Reference & 0.222 & 0.361 & 0.143 & 0.152 & 0.491 & 0.185 & 0.410 & 0.707 &0.217 & 0.425 & 0.142 \\
		\hline
	\end{tabular}
\end{table*}

\begin{table*}
	\centering
	\caption{The $M_2$$_0$ results when calculated on the segmentation map of the original target. This means only the target is considered, not background or non-target objects. We evaluate each target individually, and compare the result trends with the Gini segmentation results.}
	\label{table:M20 segmentation}
	\begin{tabular}{lccccccccccr} 
		\hline
		   & NGC1084 & NGC7222 & Ring & Dense & M53 & UGC477 & NGC4308 & NGC288 & NGC64 & NGC6543 & Virgo Cluster\\
		\hline
		 Original & -1.710 & -1.926 & -0.693 & -0.754 & -0.666 & -0.966 & -1.993 & -0.681 & -1.513 & -1.080 & -0.698\\
		 Denoised & -1.734 & -2.095 & -0.684 & -0.780 & -0.676 & -0.696 & -2.095 & -0.690 & -1.681 & -1.082 & -0.689\\
		 Reference & -1.743 & -1.992 & -0.678 & -0.931 & -0.663 & -0.896 & -2.092 & -0.685 & -1.518 & -0.986 & -0.671\\
		\hline
	\end{tabular}
\end{table*}

\begin{figure}
	\includegraphics[width=0.73\columnwidth]{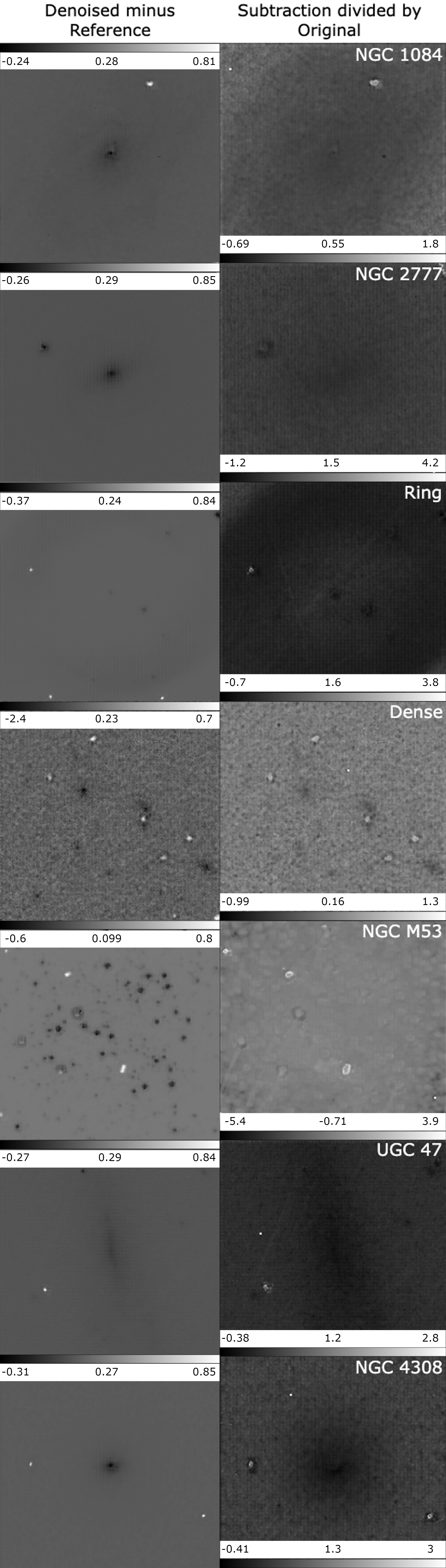}
    \caption{A collection of targets that have been created via image subtraction and division. The Denoised has been subtracted by the Reference image in order to get the resulting images in the left hand column. The right hand column is created by taking the images from the left hand column and dividing them by the Original images for each respective target. This shows an overall reduction of noise whilst not affecting the target's structure.}
    \label{fig:Subtraction Division plot Old Targets}
\end{figure}

\renewcommand{\thefigure}{\arabic{figure} (Cont.)}
\addtocounter{figure}{-1}
\begin{figure}
	\includegraphics[width=0.73\columnwidth]{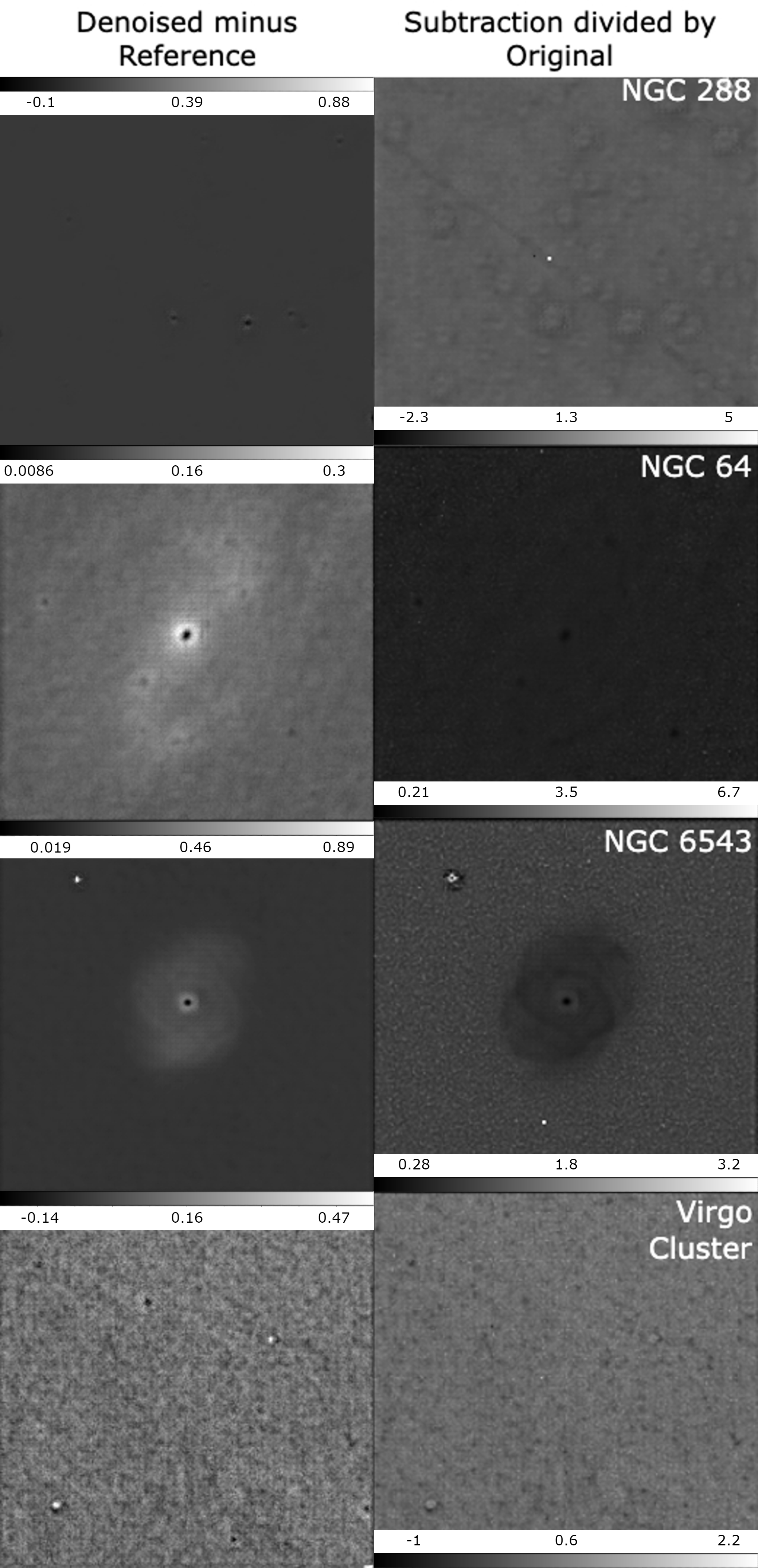}
    \caption{A collection of targets that have been created via image subtraction and division. The Denoised has been subtracted by the Reference image in order to get the resulting images in the left hand column. The right hand column is created by taking the images from the left hand column and dividing them by the Original images for each respective target. This shows an overall reduction of noise whilst not affecting the target's structure.}
    \label{fig:Subtraction Division plot New Targets}
\end{figure}
\renewcommand{\thefigure}{\arabic{figure}}

Alongside our metric results we have decided to include a subtraction and division figure to show what is being retained and what is being dismissed during our noise reduction process. Specifically we investigate into if low surface brightness features of targets are still present after the denoising process. This is seen in figure \ref{fig:Subtraction Division plot Old Targets}. We can see that for most cases the targets are preserved whilst noise across the images has been reduced. There are a few cases where the structure has been affected slightly, such as NGC 6543, but in summary the process has made a positive impact. 

\begin{figure}
\centering
\includegraphics[width=8cm]{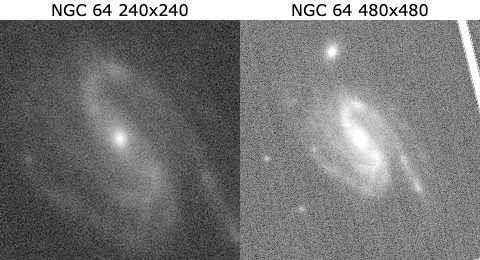}
\caption{The single-shot image (warp) of NGC 64 when the kpc scale has been changed. On the left is the version used alongside the other targets in our work, and on the right is a larger image containing more of the target, but also chip edges due to the CCD.}
\label{fig: NGC 64 240 and 480 comparison}
\end{figure}

We also investigated what the autoencoder would produce if we increased the size of the image to include more of the target. In the case of NGC 1084 for example we recognise that some of the targets morphology is cut off at the top and bottom of the image. We felt that this would be important to investigate into, but also difficult to address for the large sized galaxies that form part of the sample. As we expand the image size and increase the kpc scale to include more of the target, the images invariably encounter chip edges due to the CCD. These chip edges are very significant for single-shot images as they can dramatically influence the outcomes of metrics (e.g., $M_2$$_0$) and are problematic to remove in a clean manner without also clipping the image down. Hence why we have gone for Pan-sTARRS default 240 x 240 dimensions. With all this stated, we have managed to find a target where this issue is not as dramatic and impactful. We have investigated into NGC 64 by doubling the size of the image and re-running our approach on this version of the target that now contains more of the target within it. We report our results in table \ref{table:NGC 64 increased kpc} and show the difference between the two version in figure \ref{fig: NGC 64 240 and 480 comparison}. We note that the trends seen in the 480x480 version broadly line up with the trends seen in 240x240. Indeed, this is expected since we are not comparing absolutes, but instead the differentials, and thus clipping the sizes of images should not dramatically alter the differential comparisons made.

\begin{table*}
	\centering
	\caption{The metric results for NGC 64 when the kpc of the target image has been increased, and the input dimensions of are 480 x 480 instead of 240 x 240. Increasing the dimensions shows more of the image, but also displays chip edges of the CCD used during the survey. We also note that we keep the autoencoder settings the same, meaning a 480x480 image is downscaled in resolution to 256x256 (input layer dimensions).}
	\label{table:NGC 64 increased kpc}
	\begin{tabular}{lcccr} 
		\hline
		    & 240x240 NGC64 & 480x480 NGC64\\
		\hline
		 MSE - Original vs Noisy & 0.0029 & 0.0026\\
		 MSE - Original vs Denoised  & 0.0003 & 0.0066\\
		 MSE - Reference vs Noisy & 0.0091 & 0.0191\\
          MSE - Reference vs Denoised  & 0.0045 & 0.0052\\
          MSE - Reference vs Original & 0.0037 & 0.0194\\
          \\
		 SSIM - Original vs Noisy  & 0.62 & 0.67\\
		 SSIM - Original vs Denoised & 0.93 & 0.74\\
		 SSIM - Reference vs Noisy & 0.34 & 0.37 \\
          SSIM - Reference vs Denoised & 0.65 & 0.86\\
          SSIM - Reference vs Original & 0.64 & 0.56\\
          \\
          Gini - Original & 0.178 & 0.242\\
          Gini- Denoised & 0.185 & 0.271\\
          Gini - Reference & 0.217 & 0.355\\
          \\
          $M_2$$_0$ - Original  & -1.513 & -1.210\\
          $M_2$$_0$  - Denoised & -1.681 & -1.174\\
          $M_2$$_0$ - Reference & -1.518 & -1.151\\
		\hline
	\end{tabular}
\end{table*}

Each target has been discussed on a case-by-case basis, but it is worth nothing the general positive outcome of noise reduction, with positive signs for morphology and pixel intensity preservation on a case-by-case basis.

\section{Discussion}
\label{section:Discussion}
In terms of time the process overall takes approximately five minutes and thirty seconds. The data implementation and augmentation takes a total of one minute, and the autoencoder process takes just under four minutes. This is with the code set for developing 3000 images total, and running the autoencoder with an epoch of twenty iterations. This is also taking into consideration that we are running this on Google Colab pro, which is able to provide approximately 27 GB of RAM, operating on a Tesla T4 or P100 GPU and/or and Intel(R) Xeon(R) CPU.

To take this initial project further would mean investigating into training the network on multiple images of galaxies at once. As it stands now we train it on one image and then reset the network once we have the results. Therefore the next logical step would be for the network to retain the information of multiple galaxy formations, so that it could be used in a pipeline. With how it has performed thus far we can not see any major issue with this potential future for the work. It would also be beneficial to investigate into the removal of artefacts, and set a goal to preserve target pixel intensity more so than has been addressed in this proof of concept. Another application of this model that could coincide with the implementation of multiple galaxy formation recognition would be the use of alternate inputs. Currently we have only applied our model to images in the \emph{grizy} bands, but this can be expanded upon to take in new images from other filters and wavelengths. The next logical step then would be to supply radio inputs and see what the model makes of those. This would lead well into this model being integrated into future survey pipelines. As mentioned before, LSST and ASKAP are planned surveys due to be in full operation in the near future. Since these surveys will be observing different fields of view, different depths, and at different wavelength, this would be an ideal opportunity to expand our models capabilities and parameters whilst also establishing it in a live survey setting.  Another avenue to take with this project is to apply different target formations to the model. These could range from being more varied in order to distinguish key features amongst many different formations. Or the project could be to see if any specific feature amongst different targets of the same formation are more important than others, or have noise reduced upon them more so than other features.

\section{Conclusions}
\label{section:Conclusions}
We have managed to use a machine learning model to reduce the effect of noise on single-shot image targets. This has been achieved thanks to an autoencoder, a machine learning model. This model summaries an input image, and reconstructs the image based on the summarised data it created. Using this concept our autoencoder has been trained to identify noise on transient images and leave out as much as it can identify during the autoencoder process. This process has been applied to a variety of transient images from Pan-STARRS. The results of which have been measured by appearance to the user, as well as using measures of metrics such as SSIM, MSE and Gini. Overall we see an improvement in the denoised images compared to their original state, during which we also compare our denoised images to an alternate cleaned image created via stacking. Our application of an autoencoder has achieved these noise reduction results in a short amount of time, as well as with the input of a single transient image. This is an improvement compared to stack cleaning. 

Due to the nature of the data and how we are processing it through the autoencoder, our results cannot hope to achieve the same levels of noise reduction as previously mentioned methods. In light of this we still brand this as a new technique for noise reduction, stating that this process takes a small amount of time and processing power to achieve desirable results. And with the main feature being that it uses only one single-shot image, whereas for example stacking requires multiple, this process is one to be further implemented into ongoing survey pipelines. 

\section*{Acknowledgements}
\label{section:Acknowledgements}

OJB would like to personally thank the staff at the Milne Centre at the University of Hull for the continuous support and aid. OJB would also like to thank Student Finance England for their financial support as funder. We also would like to thank the referee at the Monthly Notices of the Royal Astronomical Society for their time and feedback on our work. 

\section*{Data Availability}
\label{section:Data availability}

Our autoencoder model is adapted from the example produced by Keras, which can be found at \href{https://keras.io/examples/vision/autoencoder/}{keras.io}. The data augmentation was inspired by \href{https://towardsdatascience.com/data-augmentation-techniques-in-python-f216ef5eed69}{towardsdatascience.com}. Our model can be found on our  \href{https://github.com/Milne-Centre/Noise-reduction-on-single-shot-images-using-an-Autoencoder}{Github} repository, wherein you can find the python code containing the image preparation, data augmentation, autoencoder model, and metrics we have used, as well as the data we have used. 

Noise calculations are derived from \href{https://mwcraig.github.io/ccd-as-book/01-03-Construction-of-an-artificial-but-realistic-image.html}{mwcraig.github.io}, and are listed in the source code for this project. 
The MSE code can be found at \href{https://pyimagesearch.com/2014/09/15/python-compare-two-images/}{pyimagesearch.com}, the SSIM code at \href{https://scikit-image.org/docs/stable/auto_examples/transform/plot_ssim.html#id1}{scikit-image.org}, and the Gini code at \href{https://github.com/oliviaguest/gini}{Github.com} with a supplementary explanation of their work found at \href{https://www.statsdirect.com/help/default.htm#nonparametric_methods/gini.htm}{statsdirect.com}.




\bibliographystyle{mnras}
\bibliography{Ref} 

\bsp	
\label{lastpage}
\end{document}